\documentclass{ws-procs9x6}

\newcommand{\M}{\mathcal{M}}
\newcommand{\hc}{^\dagger}
\newcommand{\dm}{\Delta m^2}
\newcommand{\Rs}{R_\odot}
\newcommand{\unit}[1]{\text{ #1}}
\newcommand{\dH}{\delta H}
\newcommand{\avg}[1]{\langle #1 \rangle}
\newcommand{\bavg}[1]{\bigl\langle #1 \bigr\rangle}

\begin{document}

\title{LORENTZ VIOLATION IN\\SOLAR-NEUTRINO OSCILLATIONS}

\author{JONAH E. BERNHARD}

\address{Department of Physics and Astronomy, Swarthmore College,\\
Swarthmore, PA 19081, United States\\
E-mail: jbernha2@swarthmore.edu}

\begin{abstract}
  Solar-neutrino oscillations are considered using a massive model with perturbative Lorentz violation.  The adiabatic approximation is used to calculate the
  effects of Lorentz violation to leading order.  The results are more compact than previous work involving vacuum oscillations, and are accurate for small
  Lorentz-violating coefficients.
\end{abstract}

\bodymatter

\section{Introduction}
Neutrinos are convenient for studying Lorentz violation due to their small mass and high velocities.  Previous research has considered neutrino oscillations
with short\cite{shortbaseline} and long\cite{mewes2009} baselines, but mostly zero or constant matter effects. Very little has been done with variable matter effects, such as those in the Sun.
We attempt to develop a perturbative technique for easily calculating Lorentz-violating effects in solar-neutrino oscillations.

\section{Theory}
We assume neutrinos have the standard mass differences, and treat Lorentz violation as a perturbative effect.  The neutrino Hamiltonian thus has three terms
\begin{equation}
  H = \frac{1}{2E} U\hc \dm U + V + \dH,
\end{equation}
where the first term on the right-hand side is the standard vacuum Hamiltonian, $V$ is the Sun's matter potential, and $\dH$ is a small Lorentz-violating Hamiltonian.  The diagonal
mass squared matrix $\dm$ and the unitary mixing matrix $U$ take the standard form, using the approximate observed values
\begin{align*}
  \dm_\odot &\simeq 8.0 \times 10^{-17} \unit{MeV}^2,\quad \dm_\text{atm} \simeq 2.5 \times 10^{-15} \unit{MeV}^2, \\
  \theta_{12} &\simeq 34^\circ,\quad \theta_{23} \simeq 45^\circ,\quad \theta_{13} \simeq 0^\circ,\quad \delta \simeq 0.
\end{align*}
Since $\theta_{23} \simeq 45^\circ$ and $\theta_{13} \simeq 0^\circ$, muon and tau neutrinos mix equally and $P_{e\mu} = P_{e\tau}$.  Given this similarity, it
is convenient to group muon and tau together and consider only two flavors.  In this case, there are only two parameters: a mixing angle $\theta$ and a mass difference
$\dm$, equivalent to $\theta_{12}$ and $\dm_\odot$, respectively, in the three-generation case.  This significantly simplifies many calculations.  The
two-generation case will be considered for the remainder of this work.

The Sun produces very large quantities of electron neutrinos in its core.  As they propagate to the surface, they interact with matter.  This effect can be
described by the Sun's matter potential $V$, given approximately by
\begin{equation}
   V = \begin{pmatrix}
    \sqrt 2 \, G_F \, n_e & 0 \\
    0 & 0
  \end{pmatrix},\quad
  G_Fn_e \simeq 1.32 \times 10^{-17} \, e^{-10.54R/\Rs} \unit{MeV},
\end{equation}
where $G_F$ is the Fermi coupling constant, $n_e$ is the number density of electrons in the Sun, and $\Rs$ is the solar radius.\cite{bahcall2001}

The variable matter potential precludes an exact solution; however, provided the Hamiltonian changes sufficiently slowly, approximate analytic expressions can
be obtained using the adiabatic approximation.  This approximation assumes energy eigenstates change slowly enough that neutrinos do not transition between them
as they propagate.  High-frequency oscillations between flavors can then be averaged away, resulting in simpler expressions than in previous work, such as
reference \refcite{mewes2009}.  The averages can be calculated from
\begin{equation}
  \avg{P_{eb}^{(0)}} = \bavg{|\nu_b|^2} = \sum_a \bigl| (U^*)_{ab}(U_0)_{a1} \bigr|^2
\end{equation}
where the superscript (0) indicates that these are zeroth-order (Lorentz-invariant) probabilities, $U$ is the vacuum mixing matrix, and $U_0$ is the effective
mixing matrix at $R=0$.  This evaluates to
\begin{equation}
  \avg{P_{ee}^{(0)}} = s_\theta^2 s_{\theta_0}^2 + c_\theta^2 c_{\theta_0}^2, \quad 
  \avg{P_{e\mu}^{(0)}} = s_\theta^2 c_{\theta_0}^2 + c_\theta^2 s_{\theta_0}^2,
  \label{eqn:avgP}
\end{equation}
where $s_\theta^{} \equiv \sin\theta$, $c_\theta^{} \equiv \cos\theta$, $\theta$ is the vacuum mixing angle, and $\theta_0$ is the effective mixing angle at
$R=0$.  Note that the matter potential only needs to be known at the center and surface of the Sun.

\begin{figure}
  \begin{center}
    \psfig{figure=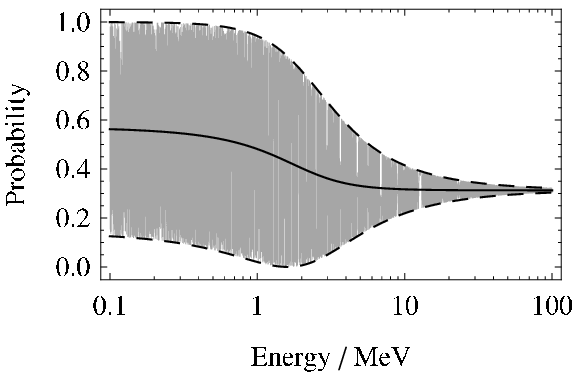,bb=0 0 165 106,clip=}
    \psfig{figure=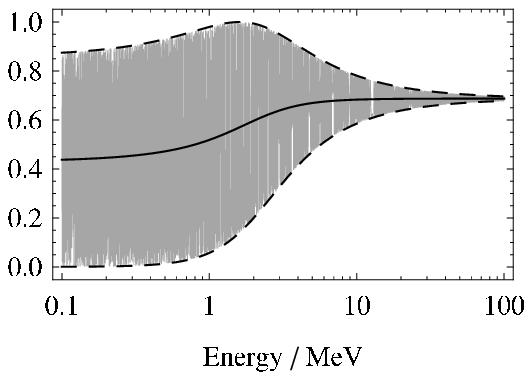,bb=10 0 165 106,clip=}
  \end{center}
  \caption{Average electron (left) and muon (right) neutrino survival probability (solid) and envelope function (dashed, given by $2s_\theta s_{\theta_0} c_\theta c_{\theta_0}$) as a function
  of energy in the adiabatic limit.  For comparison, the probability was calculated numerically at 10000 logarithmically spaced energies from 0.1 to 100 MeV
  (grey).}
  \label{fig:adiabatic}
\end{figure}

\Fref{fig:adiabatic} shows the electron neutrino survival probability as a function of energy.  The adiabatic approximation agrees with the numerical solution almost
exactly.

\section{Lorentz violation}
Lorentz violation is a natural extension to the theory.  Assuming Lorentz violation is sufficiently small, it can be treated as a first-order
perturbation to the zeroth-order probabilities from \eref{eqn:avgP}.  

The Lorentz-violating term has the form\cite{mewes2004}
\begin{equation}
  \dH_{ab} = \frac{1}{E} \bigl[ (a_L)^\alpha p_\alpha - (c_L)^{\alpha\beta} p_\alpha p_\beta \bigr]_{ab}
\end{equation}
where $(a_L)_{ab}^\alpha$, $(c_L)_{ab}^{\alpha\beta}$ are complex Lorentz-violating coefficients and $p_\alpha$ is the neutrino energy-momentum four-vector,
$p_\alpha = (E, -\vec p) \simeq E(1, -\hat p)$.  The coefficients $(a_L)_{ab}^T$, $(c_L)_{ab}^{TT}$ are isotropic and hence introduce changes to energy
dependence, while the remaining coefficients are anisotropic and cause annual variations.  The CPT-odd $(a_L)_{ab}^\alpha$ are constant with energy; the
CPT-even $(c_L)_{ab}^{\alpha\beta}$ are linear.

The probabilities from the adiabatic approximation become
\begin{equation}
  \avg{P_{eb}} = \avg{P_{eb}^{(0)}} + \frac{2E}{\dm}\,\M_{eb}^{ij}\,\dH_{ij},
\end{equation}
where $\M_{eb}^{ij}$ is a unitless, order one effective scale factor for $\dH$, used to calculate perturbed probabilities via simple linear combinations of coefficients.  
It is determined by expanding the Lorentz-violating sine and cosine to leading order, which gives
\begin{equation}
  \M_{ee}^{ij} = \dfrac{\dm}{E\,\Delta\lambda} 
  \begin{pmatrix} 
    -\frac{1}{4} \bigl( s_{2\theta}^2 c_{2\theta_0}^{} + s_{2\theta_0}^2 c_{2\theta}^{} \bigr) & 
    -s_\theta^{} c^3_\theta c_{2\theta_0}^{} - s_{\theta_0}^{} c^3_{\theta_0} c_{2\theta}^{} \\
    s^3_\theta c_\theta^{} c_{2\theta_0}^{} + s^3_{\theta_0} c_{\theta_0}^{} c_{2\theta}^{} & 
    \frac{1}{4} \bigl( s^2_{2\theta} c_{2\theta_0}^{} + s^2_{2\theta_0} c_{2\theta}^{} \bigr)
  \end{pmatrix} 
\end{equation}
and $\M_{e\mu}^{ij} = -\M_{ee}^{ij}$, where $\Delta\lambda$ is the difference between the two eigenvalues at $R=0$.

\begin{figure}
  \begin{center}
    \psfig{figure=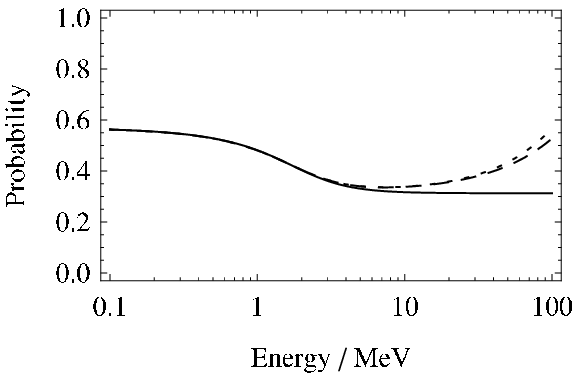,bb=0 0 165 106,clip=}
    \psfig{figure=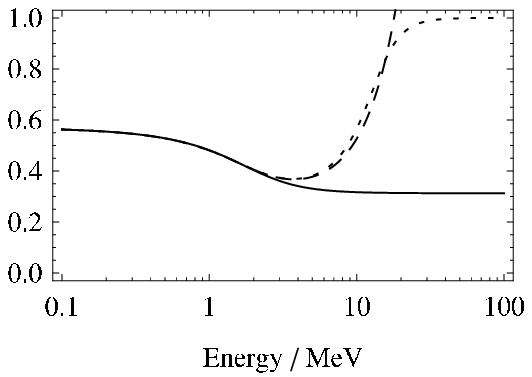,bb=10 0 165 106,clip=}
  \end{center}
  \caption{Comparison of the energy dependence of electron neutrino survival probability with no Lorentz violation (solid),  
  first-order expansion (dashed, calculated from $\M_{ee}^{ij}$), and exact (dotted, calculated numerically from the adiabatic approximation).  The
  values used are $(a_L)_{11}^T = 2 \times 10^{-19}$ MeV in the left plot, and $(c_L)_{11}^{TT} = 2 \times 10^{-19}$ on the right.}
  \label{fig:lv}
\end{figure}

\Fref{fig:lv} shows the effects of the isotropic coefficients on the energy dependence of the electron neutrino survival
probability.  The first-order expansion is effective with nonzero $(a_L)_{11}^T$, especially at low energies.  It begins to fail with nonzero $(c_L)_{11}^{TT}$
at high energies, since these effects grow with energy.  However, the Sun does not produce neutrinos with energies higher than
approximately 10 MeV, so this shortcoming is of limited relevance.

\section{Discussion}
Solar neutrinos are convenient for the study of Lorentz violation due to the large quantities and significantly lower energy and longer baseline than
accelerator experiments.  The adiabatic approximation allows simple and accurate calculations of Lorentz-violating effects in solar-neutrino oscillations.

Further work will generalize the analysis to the three-generation case and explore additional consequences of Lorentz violation, such as annual variations
caused by nonzero anisotropic coefficients and neutrino-antineutrino mixing.

\end{document}